# Towards high-rate RPC-based thermal neutron detectors using low-resistivity electrodes

**L. M. S. Margato**[a, 1], **A. Morozov**[a], **A. Blanco**[a], **P. Fonte**[a,b], **L. Lopes**[a], **J. Saraiva**[a], **K. Zeitelhack**[c], **R. Hall-Wilton**[d, e], **C. Höglund**[d, h], **L. Robinson**[d], **P. Svensson**[d], **L. Naumann**[f], **K. Roemer**[f], **D. Stach**[f], **Th. Wilpert**[g]

[a] *LIP-Coimbra, Departamento de Física, Universidade de Coimbra, Rua Larga, 3004-516 Coimbra, Portugal*

[b] *Coimbra Polytechnic–ISEC, Coimbra, Portugal*

[c] *Heinz Maier-Leibnitz Zentrum (MLZ), FRM-II, Technische Universität München, D-85748 Garching, Germany*

[d] *European Spallation Source ERIC (ESS), P.O Box 176, SE-221 00 Lund, Sweden*

[e] *Dipartimento di Fisica "G. Occhialini", Università degli Studi di Milano-Bicocca – Piazza della Scienza 3, 20126 Milano, Italy*

[f] *Helmholtz-Zentrum Dresden-Rossendorf, Bautzner Landstraße 400, 01328 Dresden, Germany*

[g] *Helmholtz-Zentrum Berlin für Materialien und Energie, Hahn-Meitner-Platz 1, 14109 Berlin, Germany*

[h] *Impact Coatings AB, Westmansgatan 29G, SE-582 16 Linköping, Sweden*

ABSTRACT:

We present experimental results on the counting rate measurements for several single-gap $^{10}$B lined resistive plate chambers ($^{10}$B-RPCs) with anodes made from standard float glass, low-resistivity glass and ceramic. The measurements were performed at the V17 monochromatic neutron beamline (3.35Å) at the Helmholtz-Zentrum Berlin. For the $^{10}$B-RPCs with 0.28 mm thick float glass a maximum counting rate density of about $8\times10^3$ Hz/cm$^2$ was obtained. In the case of low resistivity glass and ceramic, the counting rate density did not deviate from linear dependence on the neutron flux up to the maximum flux available at this beamline and exceeded a value of $3\times10^4$ Hz/cm$^2$.

Keywords: Thermal neutron detectors; Boron-10 neutron converter; Resistive plate chambers; Instrumentation for neutron sources

---

[1] *Corresponding author*: margato@coimbra.lip.pt

# 1. Introduction

We have recently shown that $^{10}$B-RPC position sensitive thermal neutron detectors (PSNDs) [1] can be designed to reach sub-millimeter spatial resolution (~0.25 mm FWHM) and high thermal neutron detection efficiency (~60% at λ=4.7Å) [2]. The RPC-based detection technology also offers a nanosecond timing resolution [3]. These properties make $^{10}$B-RPCs particularly promising for applications requiring sub-millimeter spatial resolution and TOF (time-of-flight) capability such as, e.g., TOF neutron diffraction and reflectometry [4, 5].

In our previous studies the resistive anode plates of the $^{10}$B-RPCs were made from commercial float glass (one of the most common electrode materials for RPCs), limiting the maximum counting rate density for a single-gap $^{10}$B-RPC to about $10^3$ Hz/cm$^2$ [6]. This value is below the detector requirements [5, 7, 8] for the new high flux neutron sources, such as JSNS [9], SNS [10] and ESS [11]. The counting rate capability of an RPC is mainly limited by the resistance of the electrode plates [12] and an improvement can be achieved by using a material with lower resistivity, reducing the electrode thickness or decreasing its resistivity by, for example, changing the operating temperature [12].

During the last decade several new glasses [13] and ceramics [14] were developed with resistivity in the range of $10^8 - 10^{10}$ Ω·cm. These values are significantly lower than the resistivity of the float glass, which is typically in the range $10^{12}$ to $10^{13}$ Ω·cm at room temperature. It has already been demonstrated that replacement of the float glass electrode with one made of such new materials results in an up to two orders of magnitude improvement in the counting rate capability of RPCs designed for detection of minimum ionizing particles (MIP) [13, 15, 16].

In this study we explore the benefits of using low resistivity materials in $^{10}$B-RPCs aiming to improve their counting rate capability. Experimental results on the rates achieved with single-gap $^{10}$B-RPCs with anodes made of float glass, low-resistivity glass (LR-glass) and a ceramic composite are presented. The measurements were performed with monochromatic neutrons (3.35 Å) at the V17 Detector Test Station of the Helmholtz-Zentrum Berlin (HZB).

# 2. Materials and methods

## 2.1. RPC configuration

The RPCs were assembled in a single-gap hybrid configuration (see Figure 1) with a metallic cathode lined with a $^{10}$B$_4$C neutron converter layer. Sensitivity to thermal neutrons is achieved by detection of the charged fission fragments from the neutron capture reaction $^{10}$B(n,$^4$He)$^7$Li. A detailed description of the working principles can be found in [1].

Five RPC prototypes were constructed differing only in the material of the anode, as summarized in Table 1. The ceramic plates (Si$_3$N$_4$/SiC composite) [14] were provided by Helmholtz-Zentrum Dresden-Rossendorf, Dresden, Germany. The float glass was manufactured by AGC Glass and the LR-glass is described in [13].

The cathodes (0.5 mm thick aluminium plates) were coated on the side facing the gas gap with a 1.15 μm thick layer of $^{10}$B$_4$C with an enrichment in $^{10}$B above 97% [17-19]. The $^{10}$B$_4$C deposition was performed at the ESS Detector Coatings Workshop in Linköping. On the side opposite to the gas gap, the anode plates were lined with a layer of resistive ink to uniformly distribute high voltage (HV) over the active area. A 0.35 mm diameter nylon monofilaments were used as spacers between the cathode and the anode to define a gas gap of a uniform thickness. The RPC components were hold together by fiberglass frames (FR4).

Table 1. Characteristics of the resistive materials used to manufacture the anodes of the prototypes.

| RPC designation | RPC1, RPC2 | RPC3, RPC4 | RPC5 |
|---|---|---|---|
| Material | Low resistivity glass | Float glass | Ceramic |
| Resistivity ($\Omega\cdot$cm), $\rho$ | ~ $4\times10^{10}$ | ~ $5\times10^{12}$ | ~ $2\times10^{10}$ |
| Relative permittivity, $\varepsilon_r$ | ~ 11 | ~ 8 | ~ 25 |
| Thickness (mm) | 1 | 0.28 | 2 |
| Anode area (mm $\times$ mm) | 45$\times$45 | 45$\times$45 | 80$\times$80 |
| Active\Ink area (mm $\times$ mm) | 37$\times$37 | 37$\times$37 | 70$\times$70 |

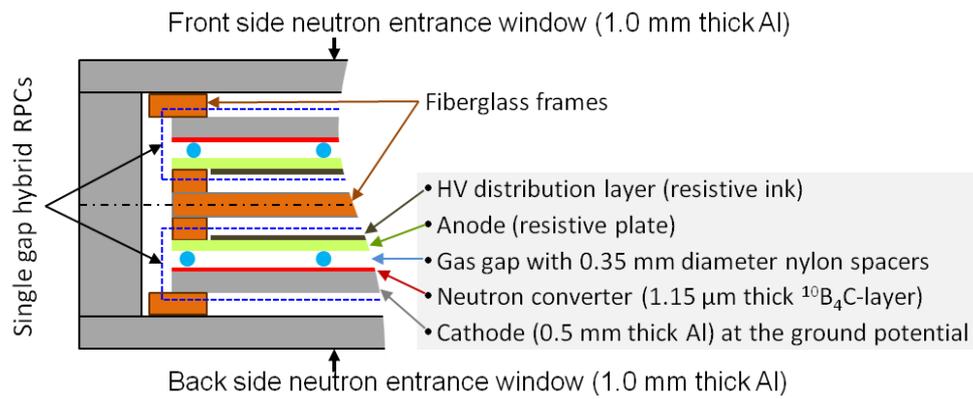

Figure 1. Schematic drawing of a cross section of the gas chamber containing the $^{10}$B-RPCs.

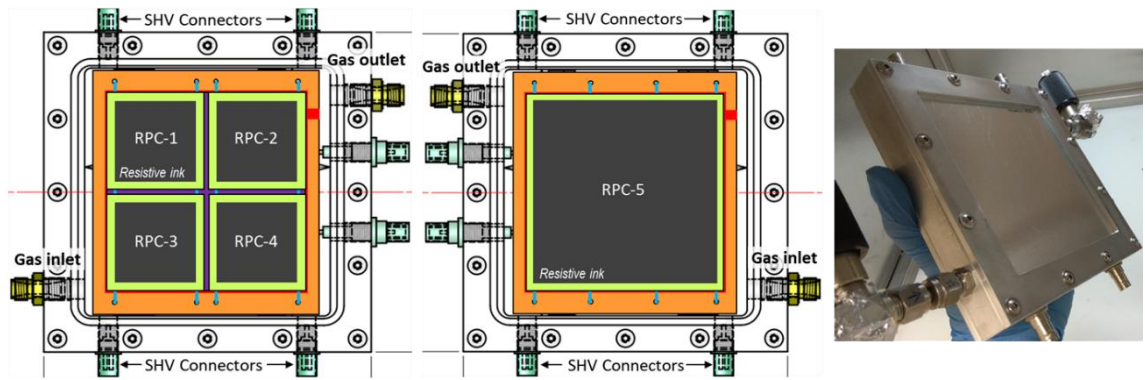

Figure 2. Schematic drawing showing the positions of the RPCs inside the gas chamber (left: front view, middle: back view) and a photograph of the gas chamber (right).

The RPCs were installed inside a gas chamber (see Figure 2). The chamber was equipped with high voltage (SHV) and signal (BNC) feedthroughs, and stainless-steel tube fittings for the gas inlet and outlet. Two 1mm thick aluminium neutron entrance windows, one in the front and the other at the back side of the chamber, were sealed with rubber O-rings.

The RPC1, 2, 3 and 4 (45×45 mm$^2$ anode area each) were assembled with a common cathode facing the front-side window. RPC1 and 2 had an LR-glass anode, and RPC3 and 4 had a float glass one. Two of each kind were made to have a back-up in case of malfunction developed during the detector transportation. The RPC5, with a ceramic anode plate (80×80 mm$^2$ area), was assembled with the cathode facing the back-side window. Each RPC could be polarized (or grounded) independently.

During the beam tests the chamber was oriented in such a way that the cathode of the operated RPC was facing the incoming neutron beam. With this arrangement the neutron beam was "seeing" only the entrance window and the cathode before interacting with the $^{10}B_4C$ neutron converter.

## 2.2. Experimental setup

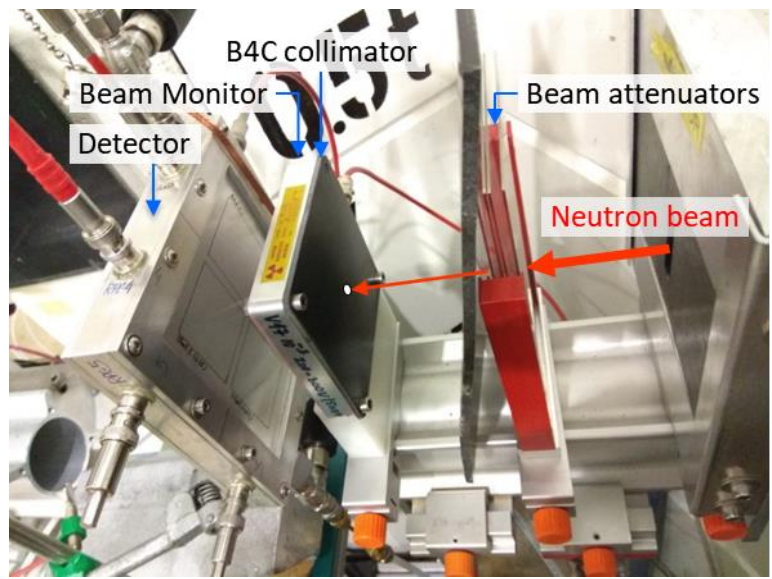

**Figure 3.** Experimental setup at the V17 detector test station at HZB. Neutron beam is attenuated by a set of glass plates and collimated by a 6 mm diameter aperture in a 3 mm thick plate of B$_4$C. After collimation the beam passes a beam monitor and irradiates the test chamber, installed on a remotely-controlled XY table.

The test chamber was installed at the V17 station at the BER II research reactor at HZB. A pyrolytic graphite monochromator (8 cm × 1 cm) provided a beam of cold neutrons with a wavelength of 3.35 Å. Figure 3 shows a photograph of the experimental setup. From right to left, the incident beam first passes a set of 1 mm thick glass attenuators, used to adjust the neutron flux. Then it is collimated by a 6 mm diameter aperture in a 3 mm thick B$_4$C plate. Next, the beam passes through a neutron beam monitor (BM) and finally enters the test chamber. To maximize the available neutron flux, this setup was assembled as close to the beam port as possible.

The BM (LND 3053) is a fission chamber with a $^{235}UO_2$ coating from LND [20]. Before the start of the experiments, it was calibrated against a $^3$He tube used as a reference detector. The BM detection efficiency is obtained from the ratio between the neutron flux measured by itself and the flux measured by the $^3$He tube, corrected from the background and beam attenuation in neutron entrance windows. The systematic uncertainty in the flux measurement with $^3$He tube is ~2.2 % (1% uncertainty in the nominal detection efficiency of the $^3$He tube and 2% uncertainty in the collimator area). A BM detection efficiency of $1.9 \times 10^{-3} \pm 4\%$ (systematic plus statistical uncertainty) is determined for 3.35 Å neutrons.

Due to the low attenuation of the beam by the BM, only at the level of few percent, it was kept in the beam path during all the experiments of this study, allowing to continuously measure the neutron flux. An attenuation value of 2.3% is obtained from a simulation with Geant4 [21], agreeing with the results reported in the literature for a similar BM from the same manufacture [22].

The RPCs were operated with tetrafluoroethane (R134a) at atmospheric pressure, with the gas continuously circulating through the chamber at a flow rate of ~2 cc/min. The environmental temperature at the V17 beamline was stabilized at 24ºC.

The RPCs were polarized by applying high voltage (HV) in the range of 1800 - 2600 V to the anode, keeping the cathode at the ground potential. Note that only one RPC at a time was polarized.

## 2.3. Counting rate measurements

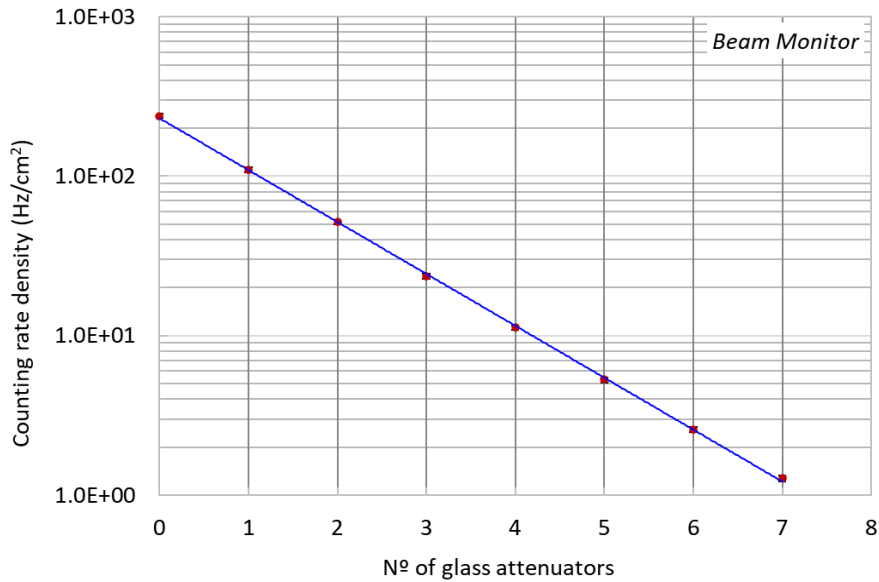

**Figure 4.** Beam monitor counting rate density as a function of the number of the glass attenuators placed in the beam path. The line is an exponential fit to the data points. The statistical uncertainties are smaller than the size of the data markers.

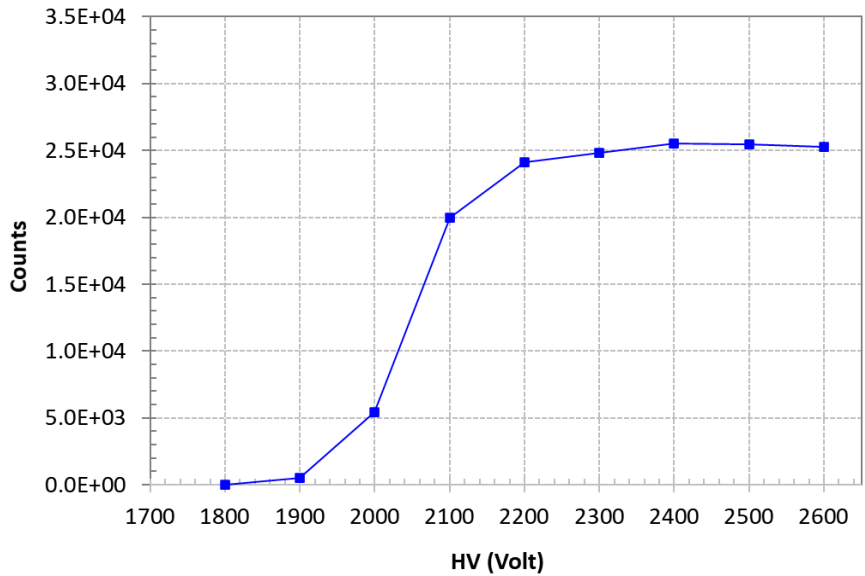

**Figure 5.** Number of counts as a function of the voltage applied to the anode, measured with the RPC3 irradiated by the neutron beam. The statistical uncertainties are smaller than the markers and the line is given just to guide the eye.

All counting rate measurements were performed using a continuous neutron beam with only a small fraction of the RPC's active area being irradiated. The test chamber was always positioned in such a way that the incident collimated beam was irradiating the polarized RPC in its geometric center.

The neutron flux was adjusted by changing the number of attenuators in the beam path. Each identical 1 mm thick glass plate attenuates the beam flux by a factor of approximately two. The linearity of the BM output with the neutron flux was confirmed over the full dynamic range of the neutron flux (see figure4).

The induced signals in the cathode were fed to a charge-sensitive preamplifier and then shaped by a linear amplifier. The amplifier's output was fed to a multichannel analyzer (MCA) and to a single-channel analyzer (SCA). The single-channel analyzer output was sent to a counter module. The detection threshold was set to an equivalent charge value in the input of the preamplifier of 60 fC.

The counting rate density was calculated as the measured number of counts per second minus the background and then divided by the collimated beam area. The background was recorded by blocking the neutron beam with a plate of boron carbide placed in front of the BM. It should be emphasized that due to constraints on the available beam time, each RPC was kept polarized only for a few hours before the start of the counting rate measurements, which is not sufficient to stabilize the dark counts. The background rates were about 0.05 Hz/cm$^2$ for the $^{10}$B-RPCs with the float glass anodes and about 0.4 Hz/cm$^2$ for the ones with the anodes in LR-glass and ceramics. The increased background in the latter case may be linked to a higher dark current due the lower bulk resistivity and differences in the surface properties of the LR-glass and ceramic electrodes, however the mechanism responsible for the dark counts in RPCs is not yet sufficiently understood [12].

## 3. Results and discussion

The counting rate was measured as a function of the incident neutron flux with the neutron beam collimated by a 6 mm diameter aperture (see section 2.2). The results for RPC3 (float glass anode) are shown in figure 6. The polarization voltage was 2300V, which is slightly above the knee of the counting plateau (figure 5). As demonstrated in figure 6, the rate is not anymore linear with the neutron flux for the counting rate densities above $4\times10^3$ Hz/cm$^2$. This maximum counting rate value is about 4 times higher than the one obtained previously by us for a similar $^{10}$B-RPC with a thicker (0.35 mm) float glass anode [6]. A similar behaviour of the counting rate capability with reduction of the electrode thickness has already been observed for float glass RPCs developed for detection of MIPs [23].

When the polarization voltage (HV) applied to the RPC3 was increased to 2500 V, the maximum counting rate density improved by a factor of two, reaching $8\times10^3$ Hz/cm$^2$ as demonstrated in figure 6. While an increase in the HV improves the maximum counting rate, it also typically leads to an increase in the dark count rate and gamma sensitivity [2]. Therefore, depending on the requirements of a particular application, the operating HV has to be chosen in order to find an acceptable compromise between the rate capability and these two parameters.

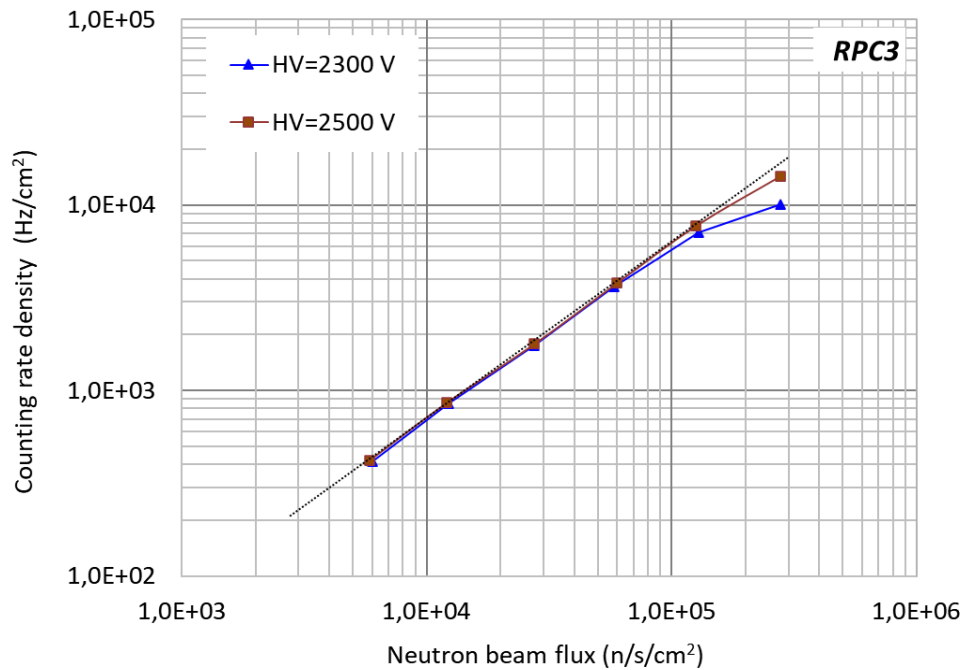

**Figure 6.** Counting rate density as a function of the neutron flux measured with RPC3 at 2300 V (triangles) and 2500 V (squares). The statistical uncertainties are smaller than the size of the markers. A thin straight line is shown to demonstrate the deviation from linear dependence.

Figure 7 shows the counting rate results for RPC1 and RPC5, which have the anode plates made from LR-glass and ceramic, respectively. The resistivity of these two materials is from two to three orders of magnitude lower than that of the float glass (see table 1). For comparison, the results for RPC3 are also depicted in the same figure. All these measurements were performed at the same polarisation voltage of 2300 V. The counting rates for RPC1 and RPC5 show no deviations from linear dependence up to the maximum neutron flux available at this time at the V17 beamline. Neutron counting rate densities exceeding $3\times10^4$ Hz/cm$^2$ were measured at these conditions.

The loss of linearity in the counting rate observed for RPC3 can be attributed to a reduction in the potential difference across the gas gap, $\Delta V_g$, developing with the increase of the rate [12, 24]. The decrease in $\Delta V_g$ results in a reduction of the gas gain, and, consequently, to a decrease in the amplitudes of the induced signals. This effect is demonstrated in figure 8 which shows the pulse height spectra (PHS) recorded with RPC3 at several neutron fluxes. There is a strong shift of the peak position toward lower amplitudes with increase of the neutron flux (reduction of the number of glass attenuators). As a result of this peak shift, a larger number of events have an amplitude below the detection threshold, leading to a decrease in the detection efficiency and, consequently, to a drop in the counting rate.

Figure 9 shows the PHS acquired for RPC5 under the same conditions. The peak position shifts significantly less with increase of the neutron flux. Comparing figures 8 and 9, one can see that using the same threshold level (vertical straight line), the most of the spectrum area stays above it for RPC5 at all irradiation conditions, in contrast to the case of RPC3, where for high fluxes a significant part of the peak area is below. Note that for the case of RPC3, the PHS data were not recorded with an unattenuated beam since the spectrum was no longer visible.

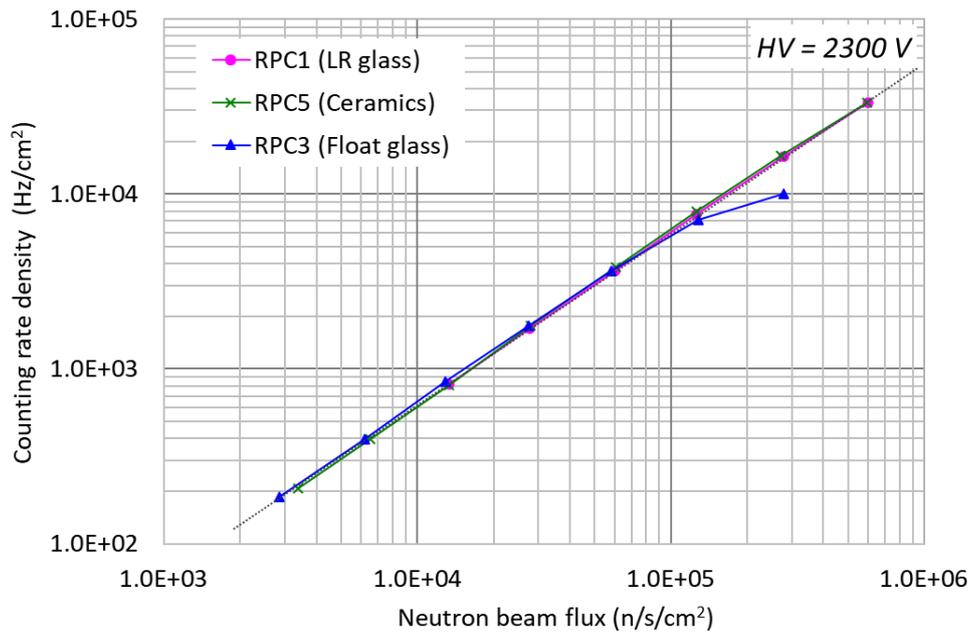

**Figure 7.** Counting rate density as a function of the neutron flux for the RPCs with the anode made from LR-glass (RPC1, round dots), ceramic (RPC5, x-shaped markers) and float glass (RPC3, triangles). All data are recorded with the polarization voltage of 2300 V. The statistical uncertainties are smaller than the size of the markers. A thin straight line is shown to demonstrate deviation from linear dependence.

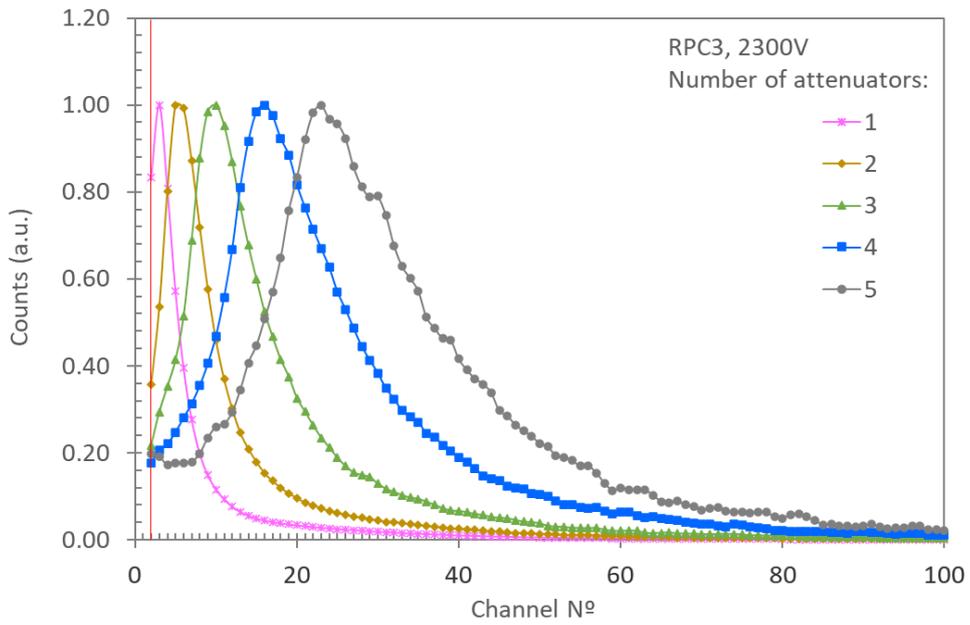

**Figure 8.** Pulse height spectra of the induced signals for RPC3 at 2300 V recorded at several neutron fluxes and normalised to the peak value. The incident neutron flux was attenuated by placing glass plates in the beam path. Each attenuator results in a reduction of about 50% in the flux.

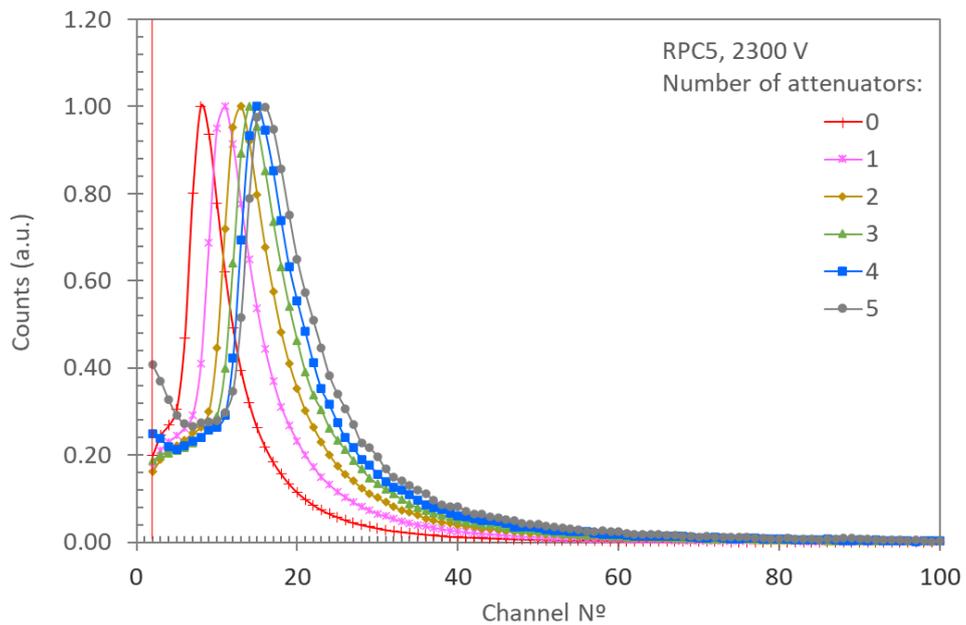

**Figure 9.** Pulse height spectra of induced signals for RPC5 at 2300 V recorded at several neutron fluxes and normalised to the peak value. The incident neutron flux was attenuated by placing glass plates in the beam path. Each attenuator results in a reduction of about 50% in the flux.

The different behaviour shown by RPC3 and RPC1/RPC5 can be explained by the difference in the bulk resistivity of their anode material. The current $I$, flowing through the resistive anode to feed the charge generated by the Townsend avalanches in the gas gap leads to development of a local drop of potential (confined to the vicinity of the avalanche) $\Delta V_r$ across the anode.

The magnitude of $\Delta V_r$ depends on the bulk resistivity $\rho$ and thickness $t$ of the resistive plate, as well as on the average charge $Q$ generated per event and on the event rate density $\phi$ (Hz/cm$^2$). Assuming a simple static ohmic model and neglecting surface conduction, $\Delta V_r$ can be expressed as [12]

$$\Delta V_r = R \cdot I = \rho \cdot t \cdot \phi \cdot Q \qquad (1)$$

where $R = \rho \cdot t/A$ and $I = Q \cdot \phi \cdot A$ ($A$ is the irradiated area). The local potential difference across the gas gap is thus given by $\Delta V_g = HV - \Delta V_r$, where $HV$ is the polarization potential applied to the RPC.

Equation 1 suggests that the resistive electrodes with lower resistivity and/or thickness result in a lower decrease in $\Delta V_g$ at the same operating conditions. This, in turn, leads to a smaller reduction in the induced charge with increase of the counting rate, thus explaining the different counting rate capabilities of RPC1/RPC5 and RPC3.

The significant improvement in the $^{10}$B-RPC counting rate capability by using low resistivity electrode materials is in line with the results reported for RPCs with that type of resistive electrodes tested with MIPs: several studies indicate an increase in the counting rate capability exceeding that of float glass RPCs by more than an order of magnitude (see e.g. [13, 16]).

Note that, the results reported here were obtained with single-gap $^{10}$B-RPCs. Another improvement in the maximum counting rate of about an order of magnitude should be possible to achieve independently of the electrode material by constructing a detector with multiple $^{10}$B-RPCs and arranging them in a multilayer or inclined-type geometry [1, 25]. In a multilayer geometry a neutron event triggers electron avalanches only in one gas gap, leaving the other undisturbed. A similar count rate improvement is not possible for RPC detectors for high energy particles (e.g., muons, protons and electrons), where an event typically triggers avalanches in multiple (possibly even all) gas gaps. Theoretically, by implementing a detector geometry with multiple $^{10}$B-RPCsand using anodes made from a low-resistivity material it should be possible to achieve local counting rates densities of several hundred kHz/cm$^2$.

## 4. Conclusions

The counting rate capabilities were characterized at a cold neutron beam (3.35 Å) for several single-gap $^{10}$B-RPCs with the anodes made from float glass and two low-resistivity materials. The resistivity of the investigated LR-glass and ceramic composite are between 2 and 3 orders of magnitude lower compared to that of the float glass.

A $^{10}$B-RPC with 0.28 mm thick float glass anodes was able to operate in linear regime with the counting rate densities up to $4 \times 10^3$ Hz/cm$^2$ at the polarisation voltage of 2300 V and up to $8 \times 10^3$ Hz/cm$^2$ at 2500 V. The $^{10}$B-RPCs with the anodes made from LR-glass and ceramic, operating at 2300 V, did not show any deviations from linear dependence up to the maximum neutron flux available at the V17 beamline of HZB. A counting rate density exceeding $3 \times 10^4$ Hz/cm$^2$ was measured at the maximum flux.

Considering these results and taking into account that an independent increase of an order of magnitude in the neutron counting rate capability can be achieved by implementing a multilayer or inclined $^{10}$B-RPC geometry [1, 25], it seems feasible to construct an RPC-based neutron detector capable to operate in linear regime with counting rate densities up to several hundred thousand counts per second per $cm^2$.

## Acknowledgments

This work was supported in part by Portuguese national funds OE, FCT-Portugal CERN/FIS-INS/0009/2019 and European Fund for Regional Development and the Program for R&D of the Sächsische Aufbaubank, Germany under the code WIDDER-100325989.

We would also like to acknowledge the Helmholtz-Zentrum Berlin for supporting the beam time at the V17 Detector Test Station.